\documentclass[conference]{IEEEtran}

\usepackage{cite}
\usepackage{amsmath,amsthm,amssymb,amsfonts}
\usepackage{graphicx}
\usepackage{textcomp}
\usepackage{xcolor}
\usepackage{color}
\usepackage{setspace}

\usepackage{mathrsfs}
\usepackage{bm}
\usepackage{pgfplots}
\usepackage{tikz}
\usetikzlibrary{arrows}
\usepackage{subfigure}
\usepackage{graphicx,booktabs,multirow}
\usepackage{upgreek}
\usepackage{bm}
\usepackage{mathrsfs}
\usepackage{algorithm}
\usepackage{algorithmicx}
\usepackage{algpseudocode}
\newcommand{\vct}[1]{{\bm #1}}
\newcommand{\mtx}[1]{{\bm #1}}

\definecolor{colorhkust}{RGB}{20,43,140}
\definecolor{colortsinghua}{RGB}{116,52,129}
\definecolor{color1}{RGB}{128,0,0}

\mathchardef\re="023C
\mathchardef\im="023D

\begin{document}

\title{Transfer Learning for Mixed-Integer Resource Allocation Problems in Wireless Networks}

%\and
%\and
%\and
%\and
%\and

\author{
   \IEEEauthorblockN{Yifei Shen$^\dagger$, Yuanming Shi$^\star$, Jun Zhang$^\dagger$, and Khaled B. Letaief$^{\dagger}$, \emph{Fellow, IEEE}}\\
  \IEEEauthorblockA{$^\dagger$Dept. of ECE, The Hong Kong University of Science and Technology, Hong Kong\\
  	$^\star$School of Information Science and Technology, ShanghaiTech University, Shanghai 201210, China\\
  	Email:{ yshenaw@connect.ust.hk, shiym@shanghaitech.edu.cn, \{eejzhang,eekhaled\}@ust.hk}}
}

\maketitle

\begin{abstract}
Effective resource allocation plays a pivotal role for performance optimization in wireless networks. Unfortunately, typical resource allocation problems are mixed-integer nonlinear programming (MINLP) problems, which are NP-hard. Machine learning based methods recently emerge as a disruptive way to obtain near-optimal performance for MINLP problems with affordable computational complexity. However, they suffer from severe performance deterioration when the network parameters change, which commonly happens in practice and can be characterized as the \emph{task mismatch} issue. In this paper, we propose a transfer learning method via self-imitation, to address this issue for effective resource allocation in wireless networks. It is based on a general ``learning to optimize'' framework for solving MINLP problems. A unique advantage of the proposed method is that it can tackle the task mismatch issue with a few additional unlabeled training samples, which is especially important when transferring to large-size problems. Numerical experiments demonstrate that with much less training time, the proposed method achieves comparable performance with the model trained from scratch with sufficient amount of labeled samples. To the best of our knowledge, this is the first work that applies transfer learning for resource allocation in wireless networks.
\end{abstract}

\section{Introduction}\label{sec:intro}
In wireless networks, effective resource allocation, e.g., power control and user scheduling, is vital for performance optimization \cite{han2008resource}. Unfortunately, typical resource allocation problems, such as subcarrier allocation \cite{wong1999multiuser}, user association \cite{zhang2017energy}, and access point selection \cite{shi2014group}, are mixed integer nonlinear programming (MINLP) problems, which are NP-hard in general. The complexity of global optimization algorithms, e.g., the branch-and-bound algorithm, is exponential. Thus, most of the studies focused on sub-optimal or heuristic algorithms, whose performance gaps to the optimal solution are difficult to control.

Machine learning recently emerges as a disruptive technology to balance the computational complexity and the performance gap for solving NP-hard problems, and has attracted lots of attention from the mathematical optimization community \cite{he2014learning,khalil2017learning}. This trend has also inspired researchers to apply machine learning based methods to solve optimization problems in wireless networks. For power control problems, which are continuous, it was proposed in \cite{sun2018learning} to accelerate the weighted minimum mean square error (WMMSE) algorithm by using the multilayer perceptron (MLP) to approximate the solution. Furthermore, unsupervised learning methods have been proposed in \cite{liang2018towards,lee2018deep} to achieve near-optimal results without knowing the optimal solution of power control. In \cite{shen2018scalable}, the network power minimization problem in Cloud-RANs, which is a MINLP problem, was investigated, and the pruning policy in the optimal branch-and-bound algorithm is learned to accelerate the runtime, while obtaining near-optimal performance.

%The proliferation of smart mobile devices has led to the exponentially growing mobile data traffic. Network densification has been recognized as a promising way to provide high-volume and diversified data services \cite{shi2018generalized}. Nevertheless, dense networks (DenseNets) impose new challenges on optimization. Typical design problems in DenseNets, such as user association \cite{zhang2017energy}, access point selection \cite{shi2014group}, computation offloading \cite{mao2016dynamic}, are usually combinatorial optimization problems, which is NP-hard in general. 

While the above attempts demonstrated the great potential of the ``learning to optimize" approach for resource allocation in wireless networks, applying them into real systems faces additional difficulties. The wireless networks are inherently dynamic, e.g., both the locations and the number of users are changing dynamically. Thus, the pre-trained machine learning model may be useless or suffer from severe performance deterioration as the network setting changes. Furthermore, even if the network setting does not change, different cells or regions face different situations, and different learning models will be needed. It is impractical to train an individual model for each cell because of the extremely long training time. These issues can be characterized as \emph{task mismatch}, i.e., the test setting is different from the trained one.

Transfer learning is a promising technique to address such task mismatch issues, as it can efficiently transfer a machine learning model trained for one scenario to a new one with little additional training and labeling effort \cite{zamir2018taskonomy,pan2010survey}. Based on the principle of transfer learning, in this paper, we propose a transfer learning method via self-imitation to address the task mismatch issue of the ``learning to branch-and-bound framework" \cite{shen2018scalable} for resource allocation in wireless networks. Specifically, we first train a policy for a network setting where abundant labeled samples are readily available. The learned policy is then blended with an exploration policy to explore the branch-and-bound tree for additional training samples in the new scenario. The best solution found is served as the training labels of the additional training samples, followed by fine-tuning the learned policy with these labels. A unique advantage of transfer learning via self-imitation is that it is able to tackle the task mismatch issue with a few additional training unlabeled samples, which is especially important when transferring to large-size problems. The numerical experiments demonstrate that the proposed method is able to achieve comparable performance with the model trained on sufficient samples for the test setting from scratch, with much less training time.
\vspace{-0.5em}
\section{An Imitation Learning Framework for Resource Allocation} \label{sec:framework}
\begin{figure*}[htb]
	\centering
	\includegraphics[width=0.7\textwidth]{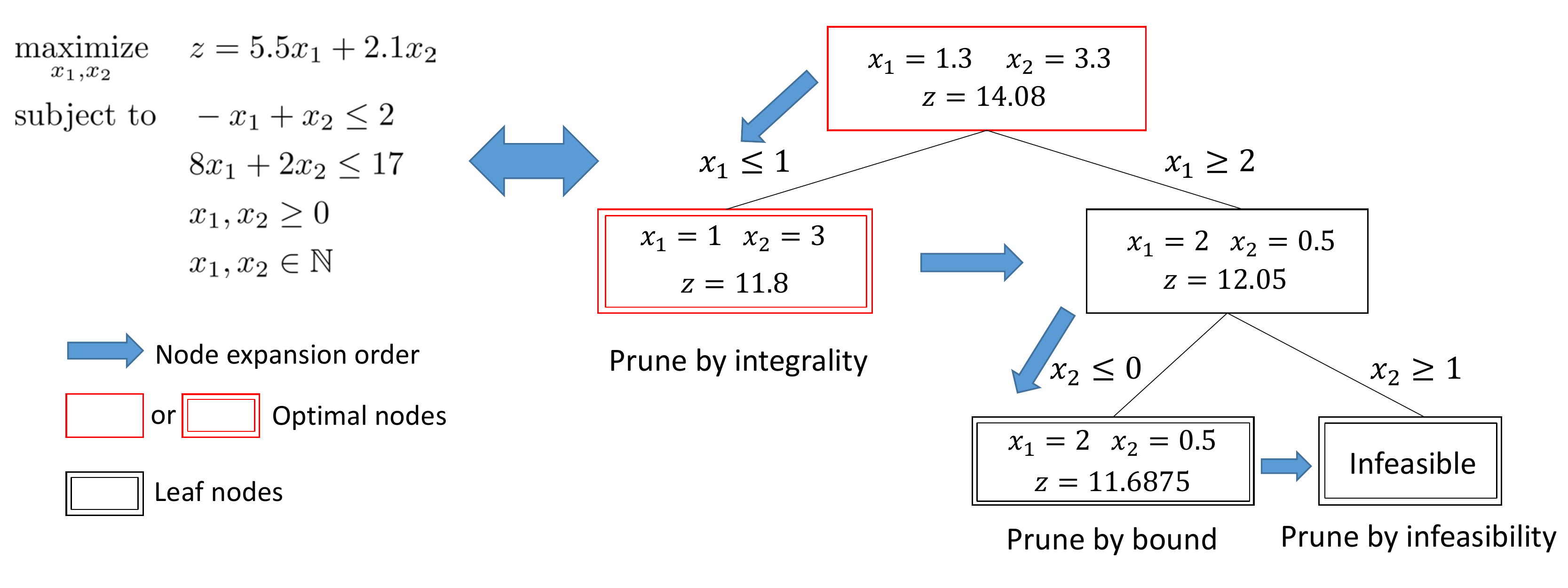}
	\caption{An example of the branch-and-bound algorithm. The solution of the relaxed linear programming problem is $x_1=1.3$, and  $x_2=3.3$. It is used as the initial solution for the tree search. The objective value $z=14.08$ provides an upper bound to the original problem. According to the branching rule, we branch out to the first variable $x_1$ and create two sub-problems. According to the node selection policy, we select the left child. The solution to the relaxed problem of the left child is $x_1=1$ $x_2=3$. Because they are integer values, we prune the node by integrality. The objective value $z=11.8$ provides a lower bound for the original problem. We then turn to the right child of the root node. We solve it and branch out on $x_2$, which generates two child nodes. The optimal objective value of the left child is $11.6875<11.8$, so we prune it by bound. The relaxed problem of the right child is infeasible and hence we prune it by infeasibility. Thus, the whole branch-and-bound tree is visited and we get the solution to the original problem $x_1=1$ $x_2=3$.}
	\label{fig:BnB}
\end{figure*}
In this section, we first present a general mixed-integer formulation for resource allocation problems in wireless networks, and then present a ``learning to optimize'' framework via imitation learning to develop near-optimal algorithms. The task mismatch issue is next identified, which motivates us to develop transfer learning based methods in the next section.

\vspace{-0.5em}
\subsection{Resource Allocation as MINLP Problems}
A wide range of resource allocation problems in wireless networks can be formulated as MINLP problems, which consist of a discrete optimization variable $\vct{a}$, e.g., indicating cell association or subcarrier allocation, and a continuous optimization variable $\vct{w}$, e.g., transmit power, subject to resource or performance constraints. Examples include network power minimization in Cloud-RAN \cite{shi2014group}, energy-efficient user association in mm-Wave networks \cite{zhang2017energy}, and subcarrier allocation \cite{wong1999multiuser}. The general formulation is given by
\begin{equation}
\begin{aligned}
&\underset{\vct{w},\vct{a}}{\text{minimize}}
& & f(\vct{a},\vct{w})\\
& \text{subject to}
& & \mathcal{Q}(\vct{a},\vct{w})\leq 0\\
&
& & a_i \in \mathbb{N}, w_i \in \mathbb{C}. \notag
\end{aligned}
\end{equation}
where $f(\cdot,\cdot)$ is an objective function, e.g., transmit power or communication delay, $a_i$ and $w_i$ are the elements of $\vct{a}$ and $\vct{w}$, and $Q(\cdot,\cdot)$ represents constraints such as the QoS or power constraint. MINLP problems are NP-hard in general. While global optimization algorithms, such as branch-and-bound, can produce optimal solutions, their computational complexity grows exponentially with the dimension of $\vct{a}$. Thus, lots of studies resort to heuristic algorithms, which, however, usually have non-negligible performance gaps compared with the globally optimal solution. To overcome these challenges, in a recent work \cite{shen2018scalable}, we proposed an effective machine learning based framework to develop efficient algorithms that achieve near-optimal solutions in such kind of problems. This framework will be presented in the next subsection.

\paragraph*{A running example} In this paper, we use the network power minimization problem in Cloud-RAN \cite{shi2014group} as a running example, as it is a typical MINLP problem. Specifically, it consists of discrete variables (i.e., the selection of RRHs
and fronthaul links), continuous variables (i.e., downlink beamforming coefficients), and the problem is a second order cone programming (SOCP) problem if the integer constraints are removed. 
\vspace{-0.5em}
\subsection{Learning to Branch-and-bound via Imitation Learning}

In this subsection, we first briefly introduce the branch-and-bound algorithm \cite{Conforti2014integer} to solve MINLP problems, followed by our learning based method to learn the pruning policy in the branch-and-bound algorithm. 

The branch-and-bound algorithm finds an optimal solution for the MINLP problem, and it consists of three main policies: the node selection policy, the variable selection policy, and the pruning policy, which together construct a binary search tree iteratively. This algorithm also maintains an unexplored node list and this list only contains the root node at the beginning. At each iteration, the node selection policy first selects a node in the unexplored list and the relaxed problem at this node is solved, by relaxing the integer constraints as continuous ones. Then the pruning policy determines whether to preserve this node according to the solution and the objective value of the relaxed problem. The algorithm enters the next iteration if the pruning policy decides to prune this node, i.e., no child node of this node will be considered. Otherwise, the variable selection policy selects a variable at this node to branch on and produce two child nodes, which are then put into the unexplored node list. 
Fig. \ref{fig:BnB} gives an example of the branch-and-bound procedure with the depth-first node selection and variable selection policy.

It is observed in the search procedure that the computational complexity is mainly determined by the pruning policy. The more nodes are pruned, the less time the branch-and-bound algorithm needs. Thus, we are able to dramatically boost the efficiency of the branch-and-bound algorithm by learning a good pruning policy. If optimality is not required, an approximate pruning policy is learned, as in \cite{shen2018scalable}. Pruning in the branch-and-bound algorithm is a \emph{sequential decision problem} because we should make a decision whether to prune a node sequentially at each iteration. There are usually two paradigms for learning the policies in sequential decision problems. The first paradigm is reinforcement learning, which has been successfully applied in many promising applications such as robot manipulation \cite{kober2013reinforcement} and Go playing \cite{silver2016mastering}. It views the features as the state and makes decisions according to the state. A reward is then given to indicate whether this decision is good or not. The goal is to maximize the reward. A requirement of employing reinforcement learning is that the reward signal should be dense \cite{songlearning}. However, for solving the MINLP problem, the learning algorithm cannot get a reward before it obtains a feasible solution, and the feasible region can be very small, which results in sparse reward signals. 

We instead turn to another paradigm, namely imitation learning \cite{daume2012course,he2014learning}. The difference between reinforcement learning and imitation learning is that there is an optimal policy in imitation learning. The imitation learning algorithm mimics the optimal policy instead of maximizing the reward. Thus, it is not influenced by the sparsity in the reward signal. In learning to prune in the branch-and-bound algorithm, the optimal policy only preserves the nodes whose feasible sets contain the optimal solution, which are referred as \emph{optimal nodes}, and prune all the remaining nodes. In the learning based framework \cite{shen2018scalable}, we first use the optimal policy to generate a label for each node, i.e., whether to prune this node or not. Then the support vector machine (SVM) is employed to learn the relationship between the features and the labels of each node. One may refer to \cite{shen2018scalable} for the detailed algorithm.

\vspace{-0.5em}
\subsection{Neural Networks as Classifiers}\label{sec:mlp}
In \cite{shen2018scalable}, SVM is used as the classifier. In this paper, we propose to use neural networks as classifiers. This will not only improve the efficiency of the framework, but also assist the developing of the transfer learning approach.

Recall that the pruning policy is a binary classification problem, where the input is the feature vector of the node and the output is a binary class in $\{prune,preserve\}$. Assuming that an $L$-layer multilayer perceptron (MLP), a type of neural networks, is used, the $k$-th layer's output is calculated as:
\begin{equation}
\vct{g}^k = \text{Relu}(\mtx{W}^k\vct{g}^{k-1}+\vct{b}^k),
\end{equation}
where $\mtx{W}^k$ and $\vct{b}^k$ are the learned parameters of the $k$-th layer. $\vct{g}^k,k=1,\cdots,L,$ denotes the output of the $k$-th layer and $\vct{g}^0$ is the input feature vector. $\text{Relu}(\cdot)$ is the rectified linear unit function, i.e., $\max(0,\cdot)$. The output indicates the probability of each class, which is a normalization of the last layer's output $\vct{g}^L \in \mathbb{R}^2$:
\begin{equation}
\vct{e}[i] = \frac{\exp(\vct{g}^L[i])}{\sum_{j=1,2}\exp(\vct{g}^L[j])},i=1,2,
\end{equation}
where $\vct{e}[i]$ indicates the $i$-th component of vector $\vct{e}$.

In the training stage, the loss function is the weighted cross entropy given by:
\begin{equation}
\ell = -\sum_{j=1,2} \vct{w}[j] \cdot \vct{y}[j]\log(\vct{e}[j])
\end{equation}
where $\vct{y}$ is the label, i.e., $\vct{y}=(1,0)$ indicates it belongs to the class \emph{pruning}, and $\vct{y}=(0,1)$ otherwise. $\vct{w}$ denotes the weight of each class, which is tuned by hand. Two parts contribute to the weight. Firstly, if the number of non-optimal nodes is much larger than the number of optimal nodes, we should assign a higher weight to the class $preserve$ in order to let the algorithm not ignore this class. We denote this part as $\vct{w}_1$ and it can be computed by:
\begin{equation}
\begin{aligned}
\vct{w}_1[1] &= \frac{\text{\# optimal nodes in the training set}}{\text{\# nodes in the training set}} \\ 
\vct{w}_1[2] &= 1 - \vct{w}_1[1].
\end{aligned}
\end{equation}
Secondly, when the number of feasible solutions is small, we should assign a higher weight to optimal nodes in the training dataset in order not to miss good solutions. This parameter is tuned by hand to achieve good performance on the validation dataset. We denote this part as $\vct{w}_2$. The total weight is calculated by $\vct{w} = \vct{w}_1 \odot \vct{w}_2$, where $\odot$ is a hadamard product.
\vspace{-0.5em}
\subsection{Task Mismatch Issue}
While the imitation learning framework demonstrated its effectiveness in \cite{shen2018scalable}, there are a few obstacles for its practical implementation. In the following, we shall elaborate on these issues. 

%\paragraph*{Dynamic Network Setting}: 
An essential assumption of most machine learning algorithms is that the training and future testing data are in the same feature space with the same distribution, i.e., the same task \cite{pan2010survey}. For example, the same task for power minimization in a Cloud-RAN means fixed numbers and locations of remote radio heads (RRHs) and mobile users (MUs), and the fixed target SINR. The performance deteriorates when the machine learning task to be applied is different from the trained one, which is referred to as the \emph{task mismatch} issue. A straightforward way to resolve this issue is to collect enough additional training samples and to train a neural network for the new setting from scratch. This will achieve good performance, but impractical in real systems because training neural networks requires large amounts of samples and long computing time. To cope with the dynamics of wireless networks, it is highly desirable to reduce the training time for the new task, i.e., when the network setting changes. Note that although the tasks are different, they share something in common, i.e., the same structure of the underlying optimization problem. In other words, the knowledge learned in the old task can be helpful for the new task. Thus, it is possible to train a new model with only a few additional samples if we can effectively transfer such knowledge. This can significantly reduce the training time and achieve good performance in the new task. Characterizing and identifying the similarities among tasks are on-going research problems in machine learning and one can refer \cite{zamir2018taskonomy} for more information.

\paragraph*{Remark} Transfer learning can also be used for accelerating the process of training a neural network from scratch. This is important if we want to apply the method to large-size wireless networks. As the training time is much less on the small-scale wireless networks, the training process can be accelerated if we first train on a small-scale network and then transfer to large-scale networks with the assistance of additional samples.

%\paragraph*{The Scalability Issue}: 
%The scalability issue arises if we want to apply the general framework to a large-size network. This is because the learning algorithms usually need more samples for problems with larger sizes. Labeling such large amounts of samples or training the neural network on such large number of samples is extremely time-consuming. As the training time is much less on the small-scale wireless networks, the scalability issue can be solved if we first train on small-scale networks and then generalize to large-scale networks with the assistance of additional samples. Thus fundamentally the scalability problem is similar to task mismatch issue, and can be solved by the same method, i.e., transfer learning. 

%In summary, the roots for these two issues are different, but both of them can be solved by transfer learning.
\vspace{-0.5em}
\section{A Self-Imitation Approach for Transfer Learning}
\begin{figure}
	\centering
	\includegraphics[width=0.4\textwidth]{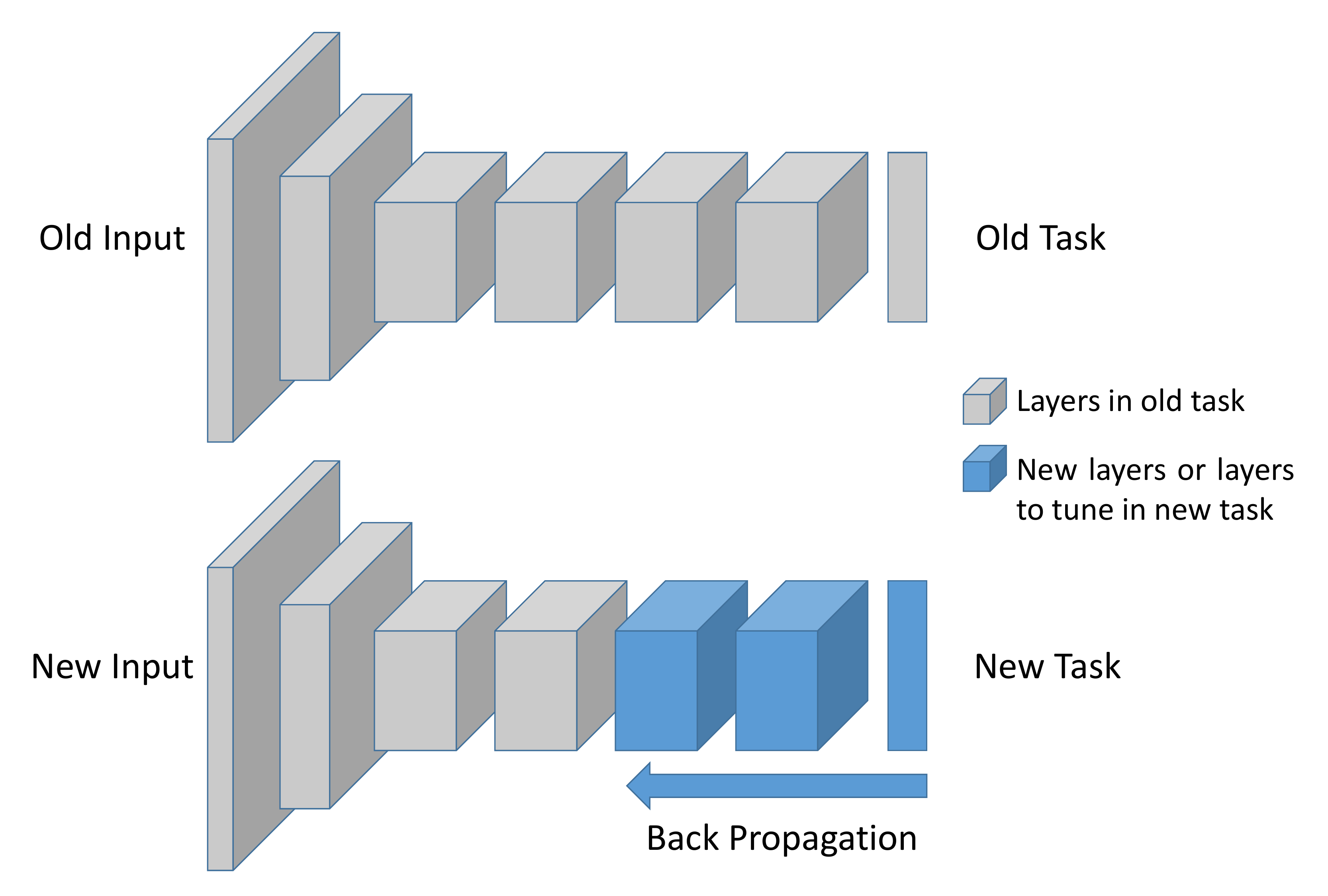}
	\caption{An illustration of transfer learning via fine-tuning. The neural network for the new task keeps the first several layers from the neural network for the old task. The last several layers are trained from scratch or tuned from those for the old task with a small learning rate.}
	\label{fig:ft}
\end{figure}
In this section, we first present a commonly used transfer learning technique in deep learning, i.e., fine tuning. Then we identify the drawbacks of this method, and present the self-imitation approach to overcome them. We employ neural networks, mainly MLP, as the machine learning method.
\vspace{-0.5em}
\subsection{Transfer Learning via Fine-tuning} \label{sec:ft}
Fine-tuning is the most frequently employed method for transfer learning in neural networks. At first, we have a neural network trained on the original training dataset, which has sufficient data and labels. We then want to apply it to a different task. The learning algorithm is highly possible not to perform well on the test dataset because of the difference in the feature distribution. Collecting sufficient samples with the same distribution as the test dataset and training a neural network from scratch is extremely time-consuming. Assuming a few samples and labels of the same task are available, we can refine the pre-trained neural network. This is the idea of transfer learning via fine-tuning. In the setting of the power minimization problem in Cloud-RAN, a task is specified by the numbers and locations of the MUs and RRHs. The tasks are different when the number of MUs or RRHs changes, but are inherently related because we are still dealing with the optimization problem of the same structure. Thus, we can apply fine-tuning, as explained in the next paragraph.

Neural networks are usually trained via stochastic gradient descent (SGD). For different layers, we can have different learning rates, i.e., the step size of the SGD. Fine-tuning is to tune the learning rate of each layer to refine the pre-trained neural network on the additional training dataset. There are two major approaches for transfer learning \cite{karpathy2016cs231n}. The first one is to set the learning rate of the first several layers as $0$. The idea is to treat the first several layers as the feature extractor, the last several layers as the classifier, and use the old feature extractor to train a new classifier. In this way, the knowledge learned by the feature extractor is transfered. The second approach is to train the pre-trained neural network with a small learning rate. The knowledge learned on the original dataset can serve as a good initialization point. The learning rate is small because we expect that the initial weights are relatively good, so distorting them too much and too quickly is not a smart choice. The illustration of fine-tuning is shown in Fig. \ref{fig:ft}.

Fine-tuning reduces the training time, but it needs additional labeled samples. The time cost is still expensive as the computational complexity of branch-and-bound to generate the labels, i.e., the optimal solutions, is exponential. This implies we even have difficulty in generating a small amount of training labels if the network size is large. Thus, it will be desirable if we can refine the model with unlabeled data, as will be proposed in the next subsection.
\vspace{-0.5em}
\subsection{Transfer Learning via Self-Imitation} \label{sec:explore}
\begin{figure*}[htb]
	\centering
	\includegraphics[width=0.8\textwidth]{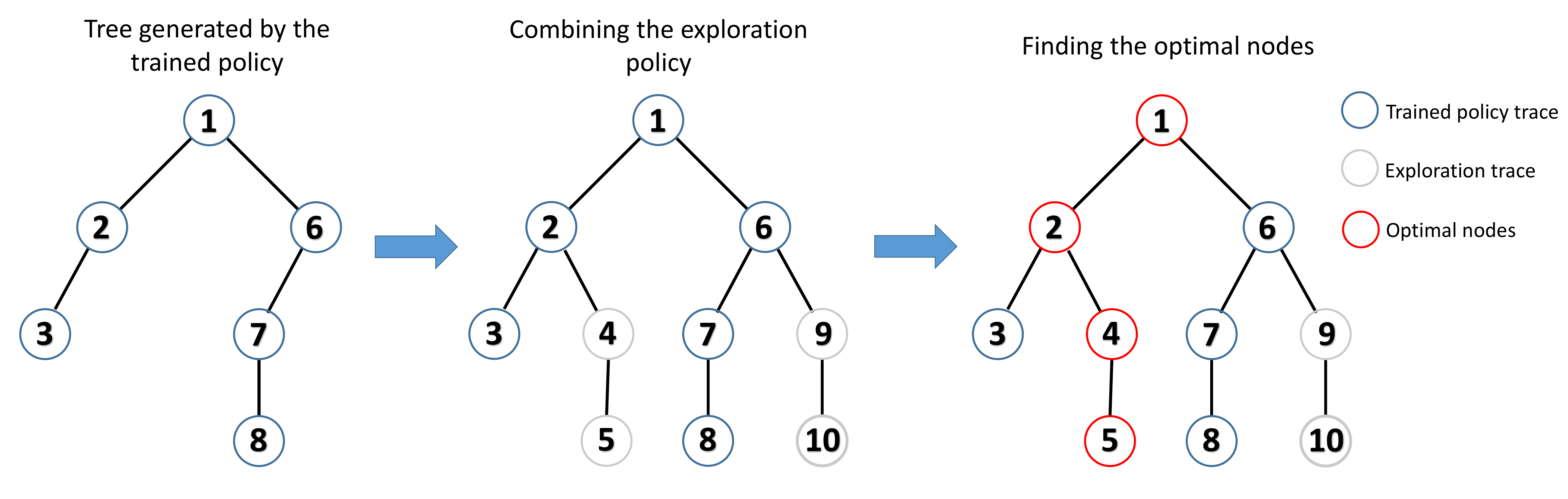}
	\caption{An illustration of the self-exploration algorithm. The number represents the order of exploration. The nodes explored in the branch-and-bound tree for the additional training samples in the new scenario include two parts: the nodes explored by the policy trained on the original dataset and the nodes explored by the exploration policy. Then the best solution found during the exploration will be used to mark optimal nodes for training neural networks. }
	\label{fig:ro}
\end{figure*}
In this section, we present an approach called \emph{self-imitation} to address the above mentioned problem for fine-tuning. We first present self-imitation learning for training the pruning policy in branch-and-bound, built upon \cite{songlearning}. Then a key component in self-imitation, i.e., the exploration policy, is described.

The main aim of self-imitation is to save the training time by avoiding generating training labels, i.e., the optimal solutions obtained by running the branch-and-bound algorithm. Since there are no training labels, the only thing the learning algorithm can access is the environment, i.e., the feedbacks from the branch-and-bound tree. This setting is similar to the reinforcement learning problem, where the agent learns the policies to control in an unknown environment by interacting with the environment. In reinforcement learning, the exploration policy is usually used to gather more information. Similarly, the self-imitation algorithm explores the branch-and-bound tree and uses the best obtained solution during the exploration as the oracle for learning.

\begin{algorithm}  
	\caption{Self-Imitation($\pi$)}  
	\label{alg:SI}  
	\begin{algorithmic}[1]
		\State $\pi^{(1)} = \pi_r$, $\mathcal{D}=\{\}$
		\For{$k = 1,\cdots,M$ }
		\For{$j = 1,\cdots,|\mathcal{P}|$ }
		\State $\hat{\pi}^{(k)} = \alpha^{(k)} \pi^{(k)} + (1-\alpha^{(k)}) \pi_e$
		\State $\mathcal{D}^{(kp)}=\text{COLLECT}(\pi^{(k)},\mathcal{P}_j)$
		\State $\mathcal{D}=\mathcal{D} \cup \mathcal{D}^{(kp)}$
		\EndFor
		\State $\pi^{(k+1)} \leftarrow$ fine-tune the classifier $\pi$ using data $\mathcal{D}$
		\EndFor
		\State \Return best $\pi^{(k)}$ on validation set
	\end{algorithmic}  
\end{algorithm}

\begin{algorithm}  
	\caption{COLLECT ($\pi$, $p$)}  
	\label{alg:COLLECT_SI}  
	\begin{algorithmic}
		\State $\mathcal{N}=\{n_0^{(p)}\},\mathcal{D}\ = \{\},i = 0,c^*=+\infty,\mathcal{N}_{opt}=\{n_0^{(p)}\}$
		\While{$\mathcal{N}\neq \emptyset$}
		\State $N \leftarrow$ select a node from $\mathcal{N}$
		\State $f \leftarrow \phi(N)$ 
		\If {$N$ is not fathomed}
		\If {$\pi(f)\neq prune$}
		\State $N_{i+1}\,N_{i+2}\ \leftarrow$ expand $N$
		\State $\mathcal{N} \leftarrow \mathcal{N} \cup \{N_{i+1},N_{i+2}\}$
		\State $i \leftarrow i+2$
		\ElsIf{$N$ contains solution to $\mathscr{P}$ and $c_N < c^*$}
		\State $c^*=c_N$
		\State $N_{o} \leftarrow N$
		\EndIf
		\EndIf
		\State $\mathcal{D} = \mathcal{D} \cup \{f\}$
		\EndWhile
		\While{$N_{o} \neq  n_0^{(p)}$}
		\State $\mathcal{N}_{opt} = \mathcal{N}_{opt} \cup N_{o}$
		\State $N_o = N_o.father$
		\EndWhile
		\For{$j \leftarrow 1,\cdots,|\mathcal{D}|$ }
		\If{$D_j.node \in \mathcal{N}_{opt}$}
		\State $\mathcal{D}_j = \{\mathcal{D}_j.f,preserve\}$
		\Else
		\State $\mathcal{D}_j = \{\mathcal{D}_j.f,prune\}$
		\EndIf
		\EndFor
		\State \Return $\mathcal{D}$
	\end{algorithmic}  
\end{algorithm}

%Learning a policy under this setting becomes a chicken-egg problem: the learning algorithm needs the right training labels to learn a good policy, but a good policy is needed to obtain those labels.

The old policy $\pi$ was trained on the original training dataset. As the task changes, in the context of Cloud-RAN, we collect some channel state information (CSI) and the fronthaul link power consumptions under the new network setting. We call it the additional training dataset, which are without labels. We blend $\pi$ with an exploration policy $\pi_e$. Then we run the DAgger \cite{ross2011reduction} with the blended policy to collect additional training samples, followed by fine-tuning the classifier on the additional training samples. The pseudo-code of the iterative training algorithm is shown as Algorithm \ref{alg:SI}. 

The data collection algorithm is different from those in the traditional imitation learning \cite{ross2011reduction}, since there is no oracle policy to produce training labels. We instead get labels by examining its past decisions. Specifically, we first perform a standard branch-and-bound procedure except using the blended policy $\hat{\pi}^{(k)}$ to prune nodes. The node features are collected without labels. We also record the best solution found $c^*$ and its corresponding node $N_o$ during the process. Then $c^*$ is used as the oracle for imitation learning. We label all the nodes on the path from the root node to $N_o$ as the optimal nodes and add these labels into the aggregation dataset $\mathcal{D}^{(kp)}$. The pseudo-code for data collection in self-imitation is shown in Algorithm \ref{alg:COLLECT_SI} and the exploration is illustrated in Fig. \ref{fig:ro}.

The key for Algorithm \ref{alg:SI} is an efficient exploration policy $\pi_e$, and its details will be provided next. A special property of learning to prune is the trade-off between the performance and the computational complexity. The more nodes are preserved, the better performance we can obtain. In Section \ref{sec:mlp}, we propose to employ the MLP to learn the pruning policy, which allows us to tune a parameter to control this trade-off during the exploration stage. 

As discussed in Section \ref{sec:mlp}, the MLP outputs $\vct{e}$, which indicates the probability of each class. During the exploration, we set a threshold to determine which class the input belongs to. In the standard classification problem, the threshold for the class $prune$ is $0.5$. If $e[1]>0.5$, the input should belong to the first class, and otherwise, it should belong to the second class. To take advantage of the trade-off between the performance and computational complexity, we set biased weights for the class $prune$, i.e., we can set a lower threshold for a higher computational efficiency or a higher threshold for better performance. The aim of the exploration policy is to achieve near-optimal results on the new dataset to serve as the labels for fine-tuning. Thus, we train the MLP on the original dataset and use it as the exploration policy by setting a high threshold.
\vspace{-0.5em}
\subsection{Implementation}
In the training stage of the self-imitation algorithm, we have to go through the dataset and solve the relaxed problem on the binary search tree at each iteration. For example, second-order cone programming (SOCP) problems need to be solved for power minimization in Cloud-RAN \cite{shi2014group}. Since solving large amounts of SOCPs is time-consuming, we propose to build a lookup table to accelerate the training process. If we encounter an SOCP that has not been solved before, we solve the problem and save the problem and its solution into the lookup table. Otherwise, we directly extract its solution from the lookup table. This method is found to be very effective throughout the simulations.

%The feature vector of a node consists of the value of branching variable, the branching variable's value in the root node's problem, the branching variable's value in the root problem, and the corresponding fronthaul link power consumption \cite{shen2018scalable}. The feature vector is in $\mathbb{R}^4$. For the multilayer perceptron
\vspace{-0.5em}
\section{Numerical Experiments}
In this section, we present numerical results to evaluate the performance of the proposed self-imitation algorithm. We first study the impact of the number of additional training samples for transfer learning, and then test the efficiency and performance of the algorithm in two different scenarios.

\paragraph*{Terminology} In the following presentation, ``original training dataset" refers to training samples for a given, small-size network, i.e., the fronthaul link power values and randomly generated channel states, whose labels are the optimal solutions obtained by the branch-and-bound algorithm.  ``Additional samples" refer to a small number of unlabeled samples for the new network setting, with different MU locations while other parameters are the same as the test dataset. ``Performance gap" means the gap compared with the objective value of the optimal solution, and ``speedup of the algorithm'' means the comparison in the running time with the original branch-and-bound algorithm.
\vspace{-0.5em}
\subsection{Influence of the Sample Size}\label{sec:n_sample}
We first show the impact of the number of additional training samples in the self-imitation algorithm. The original training dataset consists of $50$ network realizations with $5$ single-antenna MUs and $10$ 2-antenna RRHs, which are uniformly distributed in $[-1000,1000]\times[-1000,1000]$ meters. The fronthaul link power consumptions of this dataset are given by $P^c_l = 5 + l,l=1,\cdots,10$. The test dataset consists of $50$ network realizations with $15$ single-antenna MUs and $10$ 2-antenna RRHs uniformly distributed in $[-1000,1000]$ meters. The fronthaul link power consumptions are the same as the original training dataset. We set target SINR as $4$dB in all datasets. In this numerical experiment, we study how the numbers of additional samples for transfer learning will influence the performance. The speedup of the algorithm and the performance gap of the proposed algorithm are shown in Table \ref{tab:trans}, with different numbers of additional training samples. 

\begin{table}[htb]
	
	\centering  
	%\large
	%\fontsize{12}{14}
	%\selectfont  
	\caption{The speedup to branch-and-bound algorithm and the gap to the optimal objective value.} 
	\resizebox{0.45\textwidth}{!}{
		\begin{tabular}{|c|c|c|c|c|c|}  
			\hline  
			%\multirow{2}{*}{Method}&  
			%\multicolumn{3}{c|}{C}&\multicolumn{3}{c|}{ D}\cr\cline{2-7}  
			\# Samples & 2 & 5 & 10 & 20 & 50\cr\hline 
			Speedup & 9.27x&9.73x &12.73x &13.2x & 13.3x \cr\hline
			Gap&0.57\%&0.56\%&0.56\%&0.57\%&0.53\%
			\cr\hline
	\end{tabular}}
	\label{tab:trans}
\end{table}

It is shown in Table \ref{tab:trans} that the self-imitation learning algorithm achieves 9.27x speedup to the traditional branch-and-bound algorithm and $0.57\%$ in performance gap with only $2$ additional samples, which demonstrates the effectiveness of this algorithm. As shown in \cite{shen2018scalable}, to directly train for the test setting from scratch, at least $50$ \emph{labeled} samples are needed. In Table \ref{tab:trans}, we see that the self-imitation algorithm achieves comparable results with training from scratch (which is shown in Table \ref{tab:sp}) in both achievable speedup and performance, with $20$ unlabeled samples. Thus, the transfer learning based method needs fewer samples, without labels.

\vspace{-0.5em}
\subsection{Performance Evaluation}\label{sec:per}
In this section, we evaluate the performance of the self-imitation algorithm via two numerical experiments. In the first experiment, the numbers of MUs in the original training dataset and test dataset are different, which commonly happens in practice. In the second experiment, the original training dataset differs from the test dataset in both the numbers of MUs and RRHs. This situation happens when we train a model for one area/network and want to apply it to another area/network. In order to compare the performance of the proposed algorithm under these two different settings, we use the same test dataset and additional training dataset for the two experiments. The test dataset consists of $50$ network realizations with $10$ 2-antenna RRHs and $15$ single-antenna MUs uniformly distributed in the square region $[-1000,1000] \times [-1000,1000]$ meters. The fronthaul link power consumptions are given by $P^c_l = (5+l)W,l=1,\cdots,10$. The self-imitation algorithm is transferred on the additional training dataset, which consists of $20$ additional samples.

The reference method is trained on $50$ network realizations with the same task as the test dataset, i.e., training from scratch. Its performance gap is shown in Table \ref{tab:gap}, and its running time on the test dataset compared to the branch-and-bound algorithm is shown in Table \ref{tab:sp} as ``Train From Scratch".

\begin{table}[htb]
	
	\centering  
	%\large
	%\fontsize{12}{14}
	%\selectfont  
	\caption{The gap to the optimal objective value.} 
	\resizebox{0.45\textwidth}{!}{
		\begin{tabular}{|c|c|c|c|c|c|}  
			\hline  
			%\multirow{2}{*}{Method}&  
			%\multicolumn{3}{c|}{C}&\multicolumn{3}{c|}{ D}\cr\cline{2-7}  
			Target SINR & 0 & 1 & 2 & 3 & 4\cr\hline
			{\tiny Train From Scratch}&0.8\%&2.3\%&0.6\%&1.7\%&0.7\%\cr\hline 
			{\tiny Transfer For Dynamic MUs}&0.7\%&2.3\%&0.5\%&1.6\%&0.6\%\cr\hline
			{\tiny Transfer For Different Networks}&0.7\%&2.4\%&0.8\%&1.9\%&0.5\%\cr\hline  
	\end{tabular}}
	\label{tab:gap}
\end{table}

\begin{table}[htb]
	
	\centering  
	\caption{The speedup of the algorithms to the branch-and-bound.} 
	%\large
	%\fontsize{12}{14}
	%\selectfont  
	
	\resizebox{0.45\textwidth}{!}{
		\begin{tabular}{|c|c|c|c|c|c|}  
			\hline  
			%\multirow{2}{*}{Method}&  
			%\multicolumn{3}{c|}{C}&\multicolumn{3}{c|}{ D}\cr\cline{2-7}  
			Target SINR & 0 & 1 & 2 & 3 & 4\cr\hline
			{\tiny Train From Scratch}&25.2x&16.3x&23.0x&15.4x&13.2x\cr\hline 
			{\tiny Transfer For Dynamic MUs}&22.9x&16.5x&23.2x&12.8x&13.3x\cr\hline 
			{\tiny Transfer For Different Networks}&23.4x&16.8x&24.4x&13.4x&13.4x\cr\hline 
	\end{tabular}}
	\label{tab:sp}
\end{table}

\begin{table}[htb]
	
	\centering  
	%\large
	%\fontsize{12}{14}
	%\selectfont  
	\caption{Training speedup compared to ``Train From Scratch".} 
	\resizebox{0.45\textwidth}{!}{
		\begin{tabular}{|c|c|c|c|c|c|}  
			\hline  
			%\multirow{2}{*}{Method}&  
			%\multicolumn{3}{c|}{C}&\multicolumn{3}{c|}{ D}\cr\cline{2-7}  
			Target SINR & 0 & 1 & 2 & 3 & 4\cr\hline 
			{\tiny Transfer For Dynamic MUs}&11.9x&12.7x&15.6x&10.3x&10.8x \cr\hline 
			{\tiny Transfer For Different Networks}&10.1x&10.4x&14.32x&10.08x&9.65x \cr\hline 
	\end{tabular}}
	\label{tab:train_sp}
\end{table}

In the first experiment, the original training dataset consists of $50$ network realizations with $7$ single-antenna MUs and $10$ 2-antenna RRHs uniformly distributed in $[-1000,1000] \times [-1000,1000]$ meters and fronthaul link power consumptions $P^c_l = (5+l)W,l=1,\cdots,10$. For the exploration policy, we set the threshold for \emph{pruning} as $0.9$. The blended ratio is $\alpha = \text{min}(1,0.2k),k=1,\cdots,10$. Its performance gap is shown in Table \ref{tab:gap}, and its running time on the test dataset compared to the branch-and-bound algorithm is shown in Table \ref{tab:sp} as ``Transfer For Dynamic MUs". We also compute the training speedup of this method compared with the ``Train From Scratch" method in Table \ref{tab:train_sp}. The training time of ``Train From Scratch" consists of time for generating labels and training. The training time of ``Transfer For Dynamic MUs" is only composed of the time to execute the self-imitation algorithm since there are no labels required.

%The original training dataset consists of $50$ network realizations with $L=10$ 2-antenna RRHs and $K = 7$ single-antenna MUs uniformly and independently distributed in the square region $[-1000,1000] \times [-1000,1000]$. The fronthaul link power consumption is set to $P^c_l = (5+l)W, l=1,\cdots,10$. Our pre-trained network is trained on this dataset. We generate $20$ extra network realizations of the same number of MUs with the test dataset and perform the self-imitation algorithm on it. The performance and speedup of self-imitation are shown in Table \ref{tab:user_change}. For exploration policy, we set the threshold for \emph{pruning} as $0.9$. The blended ratio $\alpha = \text{min}(1,0.2k),k=1,\cdots,10$. Its performance gap to the optimal solution is shown in Table \ref{tab:user_change} and its running time on the test dataset compared to the branch-and-bound algorithm is shown in Table \ref{tab:comp} as ``MU Transfer". 

%For fine-tuning, we generate $10$ extra network realizations of the same number of MUs with the test dataset. We use branch-and-bound to get the optimal solution of them and fine-tune the pre-trained neural network on it. For self-imitation, 

In the second experiment, the original training dataset consists of $200$ network realizations with $6$ 2-antenna RRHs and $6$ single-antenna MUs uniformly and independently distributed in the square region $[-1000,1000] \times [-1000,1000]$. The fronthaul link power consumptions are uniformly distributed in $[6,15]W$. The performance gap, the speedup in the running time of the trained model to the branch-and-bound algorithm, and the training speedup are shown in Table \ref{tab:gap}, Table \ref{tab:sp}, and Table \ref{tab:train_sp}, respectively, labeled as ``Transfer For Different Networks".

As shown in Table \ref{tab:gap} and Table \ref{tab:sp}, under two different settings, the self-imitation algorithm achieves comparable performance with the reference method, in both speedup and performance gap. Additionally, we see that the self-imitation algorithm speeds up the training process by a factor of about $9$ to $15$ in Table \ref{tab:train_sp}. This indicates that the self-imitation algorithm enables the learning to branch-and-bound framework to adapt to the new setting with much less training time. The training speedup arises for two reasons. Firstly, it transfers the knowledge learned in the task of the original dataset to the task of the test dataset, and thus it requires fewer samples. Secondly, the self-imitation algorithm requires no additional training labels and thus the time for generating labels can be saved. Besides, ``Transfer For Dynamic MUs" has a slightly higher training speedup compared to ``Transfer For Different Networks". This is because the original training dataset and the test dataset are more similar in ``Transfer For Dynamic MUs" than those in ``Transfer For Different Networks".

\section{Conclusions}
In this paper, we presented a transfer learning method via self-imitation algorithm to address the task mismatch issue when applying machine learning based methods for resource allocation in wireless networks. A unique advantage of this method is that it only needs a minimal number of extra samples without training labels for transfer learning, and thus significantly reduces the training time. This is the first attempt in applying transfer learning to resource allocation in wireless networks, and it is interesting to test the effectiveness of the proposed approach or fine-tuning to other machine learning based methods for wireless networks.

%\newpage
\vspace{-0.5em}
\bibliographystyle{ieeetr}
\bibliography{ref}

\end{document}